\RequirePackage{snapshot}
\documentclass[showpacs,aps,prd,nofootinbib,floatfix,amsmath,amssymb,10pt,twocolumn]{revtex4-2}
\usepackage{graphicx}
\usepackage[utf8x]{inputenc}
\usepackage{color}
\usepackage{dsfont}
\usepackage{subfig}
\usepackage{multirow}
\usepackage{hyperref}
\usepackage{slashed}
\usepackage{cancel}
\usepackage{pdfsync}
\usepackage{booktabs}
\usepackage{float}
\usepackage{natbib}

\begin{document}

\title{Flavor-changing top quark rare decays in the Bestest Little Higgs Model}

\author{T. Cisneros-P\'{e}rez}
\email{tzihue@gmail.com}
\affiliation{\small Unidad Acad\'emica de Ciencias Qu\'{\i}micas, Universidad Aut\'onoma de Zacatecas\\
Apartado Postal C-585, 98060 Zacatecas, M\'exico.}

\author{ M. A. Hern\'andez-Ru\'{\i}z
}
\email{maria.hernandez@uaz.edu.mx}
\affiliation{\small Unidad Acad\'emica de Ciencias Qu\'{\i}micas, Universidad Aut\'onoma de Zacatecas\\
Apartado Postal C-585, 98060 Zacatecas, M\'exico.}

\author{ A. Guti\'errez-Rodr\'{\i}guez}
\email{alexgu@fisica.uaz.edu.mx}
\affiliation{\small Facultad de F\'{\i}sica, Universidad Aut\'onoma de Zacatecas\\
Apartado Postal C-580, 98060 Zacatecas, M\'exico.}

\author{E. Cruz-Albaro}
\email{elicruzalbaro88@gmail.com}
\affiliation{\small Facultad de F\'{\i}sica, Universidad Aut\'onoma de Zacatecas\\
Apartado Postal C-580, 98060 Zacatecas, M\'exico.}

\begin{abstract}
 This paper investigates the effects of parameters in the Bestest Little Higgs Model (BLHM) on rare flavor-changing decays of the top quark. As a result, flavor-changing phenomena are introduced in the BLHM for the first time. In this study, we incorporate new flavor mixing terms between the light quarks of the Standard Model (SM) and the fermions and bosons of the BLHM. We compute the one-loop contributions from the heavy quark $(B)$ and the heavy bosons $(W^{\prime\pm}, \phi^{\pm}, \eta^{\pm},H^{\pm})$. Our findings demonstrate that the branching ratios of decays $t\to qV$ and $t\to qh^0$, where $q=u,c$ and $V=Z, \gamma, g $, exhibit improvements compared to their counterparts in the SM, except for the gluon case. Moreover, we observe that the processes with higher sensitivity are $Br(t\to cZ)\sim 10^{-5}$, $Br(t\to c\gamma)\sim 10^{-6}$ and $Br(t \to ch^0) \sim 10^{-8}$ within the appropriate parameter space.
\end{abstract}

\maketitle

%
%

\section{Introduction}

Being the heaviest elementary particle in nature, the top quark gives us the best chances to discover evidence of new physics Beyond the Standard Model (BSM). On the other hand, Electroweak Symmetry Breaking (EWSB) is a mechanism adaptable to various extensions of the SM that provides us with fields whose properties characterize the physics of these models. These ingredients are the primary theoretical resources in different extensions of the SM to propose new physics~\cite{arkani2002littlest,Gutierrez-Rodriguez:2014vaa,blanke2012deltaf,PhysRevD.72.034018,drees1995implications,li1994rare,susyinduced}.
Little Higgs-type models generalize these resources as they were constructed to address the hierarchy problem of the SM by implementing a global symmetry that is collectively broken at an energy scale higher than the electroweak scale~\cite{schmaltz2005little}. In the case of the BLHM~\cite{schmaltz2010bestest},
the main objective is to preserve the advantages and address the issues encountered by its predecessors. In this regard, the BLHM resolves the Higgs stability problem more effectively and naturally by introducing new bosons and symmetries to control the radiative contributions that affect this sector. It also offers a richer and more diverse phenomenology, allowing for the observation of more distinctive experimental signals and providing a greater capacity to discriminate between different BSM. However, it has been explored very little, initially only investigating the most favorable parameter spaces~\cite{kalyniak2015constraining,Godfrey:2012tf}. Currently, research in the BLHM has been intensified by calculating its effects on some observables.
The reported results have yielded good approximations to experimental data in the chromomagnetic dipole moment of the top quark $(\hat{\mu}_t \sim -10^{-4})$ \cite{Aranda:2021kza}, the sensitivity of the weak dipole moments of the top quark $(a^W_t \sim 10^{-4})$ \cite{Cruz-Albaro:2022kty}, the electromagnetic and weak dipole moments of the $\tau$ lepton $(a_{\tau} \sim 10^{-10}, a^W_{\tau} \sim 10^{-9})$ \cite{Cruz-Albaro:2022lks}, and the electromagnetic and weak dipole moments of the top quark $(a_{t} \sim 10^{-4}$, $a^W_{t} \sim 10^{-5})$ \cite{Cruz-Albaro:2023pah}.
In the case of Flavor Changing Neutral Currents (FCNC), there have been studies primarily in the Little Higgs model with T-parity (LHT). In Ref. \cite{blanke2006particle}, they calculate some observables with particle-antiparticle mixing in the LHT for the first time. In \cite{hong2007flavor}, we find the first study on rare decays of the top quark with significant results such as $Br(t\to cg)\sim10^{-2}$, $Br(t\to cZ)\sim10^{-5}$, and $Br(t\to c\gamma)\sim10^{-7}$. The decays of $K$, $B$, $D$ mesons are thoroughly examined in \cite{blanke2009fcnc,blanke2007rare,blanke2009fcnc,Blanke2016} with various parametrizations of the Cabibbo-Kobayashi-Maskawa (CKM) \cite{kobayashi1973cp} extended matrices in the LHT model. In the article \cite{han2008higgs}, they investigate the rare decays of the Higgs boson and the Z boson with interesting results such as $Br(Z\to b\bar{s})\sim10^{-7}$. In Ref. \cite{han2016revisiting}, the authors calculate the branching ratios of the decays $t\to cX$ and $t\to cXX$, where $(X=Z,\gamma, g,h)$, with the results $Br(t\to cX)\sim10^{-2}-10^{-5}$ and $Br(t\to cXX)\sim10^{-3}-10^{-8}$. All these results improved upon their counterparts in the SM by several orders of magnitude.

Our research addresses the rare decays $t\to qV$ and $t\to qh^0$, where $q=u,c$ and $V=Z, \gamma, g$  within the theoretical framework of BLHM by calculating their respective branching ratios. Through this, we quantify the contributions of heavy quarks and heavy bosons in flavor-violating processes within a model that improves upon previous LH models. Similarly, it allows us to constrain new parameters in BLHM, such as angles and phases of proposed CKM-like matrices and their potential use in other flavor and CP-violation studies.

The article is structured as follows: In Section \ref{sec2}, we introduce the BLHM, covering the gauge boson sector, the fermionic sector, and flavor mixing within the model. Section \ref{parametros} is dedicated to constructing the parameter space utilized in our study. Moving on to Section \ref{sp}, we delve into the phenomenology of flavor-changing rare decays of the top quark within the framework of the BLHM, employing various extended CKM matrices. Lastly, we present our conclusions in Section \ref{conclusiones}. Appendix \ref{feynrules} includes the Feynman rules for flavor mixing in the BLHM.

%
%

\section{Review of the BLHM}
\label{sec2}

The BLHM~\cite{schmaltz2010bestest} arises from a symmetry group $SO(6)_A\times SO(6)_B$ that breaks at the scale $f$ towards $SO(6)_V$ when the non-linear sigma field $\Sigma$ acquires a vacuum expectation value (VEV), $\langle\Sigma\rangle=1$. There are now 15 pseudo-Nambu Goldstone bosons parameterized by the electroweak triplet $\phi^a$ $(a=1,2,3)$ with zero hypercharges and the triplet $\eta^a$ where $(\eta_1,\eta_2)$ form a complex singlet with hypercharge and $\eta_3$ is a real singlet,
\begin{equation}\label{Sigma}
\Sigma=e^{i\Pi/f}e^{2i\Pi_h/f}e^{i\Pi/f},
\end{equation}

\begin{equation}
\Pi=\begin{pmatrix}
\phi_a T^a_L+\eta_aT^a_R & 0 & 0\\
0 & 0 & i\sigma/\sqrt{2}\\
0 & -i\sigma/\sqrt{2} & 0
\end{pmatrix},
\end{equation}
\begin{equation}
\Pi_h=\begin{pmatrix}
0_{4\times4} & h_1 & h_2\\
-h^T_1 & 0 & 0\\
-h^T_2 & 0 & 0
\end{pmatrix},
\end{equation}
where $h_i^T=(h_{i1},h_{i2},h_{i3},h_{i4})$, $(i=1,2)$, are Higgs \textbf{4}'s of $SO(4)$. $\sigma$ is a real scalar field necessary to produce a collective quartic coupling~\cite{schmaltz2010bestest}. $T_{L,R}^a$ are the generators of $SU(2)_L$ and $SU(2)_R$.

%
%

\subsection{Scalar sector}
In the BLHM, two operators are necessary for the quartic coupling of the Higgs to be generated under collective symmetry breaking, but none of the operators alone allows the Higgs to acquire a potential.
In this way, we can write this potential quartic as~\cite{schmaltz2010bestest}
\begin{equation}\label{potVq}
 V_q=V_{q_1}(f,\Sigma)+V_{q_2}(f,\Sigma).
\end{equation}

$V_{q_1}$ breaks $SO(6)_A\times SO(6)_B\to SO(5)_{A5}\times SO(5)_{B6}$, where $SO(5)_{A5}$ prevents $h_1$ from acquiring a potential, and $SO(5)_{B6}$ does the same for $h_2$. $V_{q_2}$ breaks $SO(6)_A\times SO(6)_B\to SO(5)_{A6}\times SO(5)_{B5}$. If we expand Eq. (\ref{Sigma}) in powers of $1/f$ and substitute into Eq. (\ref{potVq}), we obtain
\begin{equation}\label{potVqSerie}
 V_q=k_{\pm}\left(f\sigma\pm\frac{1}{\sqrt{2}}h_1^Th_2+\dots\right)^2,
\end{equation}
where $k_{+}=\lambda_{56}/2$ and $k_{-}=\lambda_{65}/2$ are two constants. From Eq. (\ref{potVqSerie}), a mass term for scalar $\sigma$ is generated,
\begin{equation}
 m_{\sigma}^2=(\lambda_{56}+\lambda_{65})f^2.
\end{equation}
If we do not consider gauge interactions, not all scalars acquire mass, so we must add the potential
\begin{equation}\label{potVs}
 V_s=V_{s_1}(f,m_4,\Delta,\Sigma)+V_{s_2}(f,m_5\Sigma_5,m_6\Sigma_6),
\end{equation}

\noindent where $m_4$, $m_5$, and $m_6$ are mass parameters, and $(\Sigma_5,\Sigma_6)$ are matrix elements of Eq. (\ref{Sigma}). The $\Delta$ operator, comes from a global symmetry $SU(2)_C\times SU(2)_D$ that is broken to a diagonal $SU(2)$ at the scale $F>f$ when develops a VEV, $\langle\Delta\rangle=1$. We can parameterize it in the form
\begin{equation}
 \Delta=e^{2i\Pi_d/F},\hspace{0.3cm}\Pi_d=\chi_a\frac{\tau_a}{2}\hspace{0.3cm}(a=1,2,3),
\end{equation}
  where matrix $\Pi_d$ contains the scalars of the triplet $\chi_a$ that mix with the triplet $\phi_a$, and $\tau_a$ are the Pauli matrices. $\Delta$ is connected with $\Sigma$ in such a way that the diagonal subgroup of $SU(2)_A\times SU(2)_B\subset SO(6)_A\times SO(6)_B$ is identified as the SM $SU(2)_L$ group.
  If we expand the operator $\Delta$ in powers of $1/F$ and substitute it into Eq. (\ref{potVs}), we obtain
\begin{equation}
 Vs=\frac{1}{2}\left(m^2_{\phi}\phi^2_a+m^2_{\eta}\eta^2_a+m^2_1h^T_1h_1+m^2_2h^T_2h_2\right),
\end{equation}
where
\begin{eqnarray}
 m^2_{\phi}&=&m^2_{\eta}=m^2_4,\\\nonumber
 m^2_1&=&\frac{1}{2}(m^2_4+m^2_5),\\\nonumber
 m^2_2&=&\frac{1}{2}(m^2_4+m^2_6).
\end{eqnarray}
To trigger EWSB, the next potential term is introduced\cite{schmaltz2010bestest}:
\begin{equation}
 V_{B_{\mu}}=m_{56}^2f^2\Sigma_{56}+m_{65}^2f^2\Sigma_{65},
\end{equation}
where the mass terms $m_{56}$ and $m_{65}$ correspond to the matrix elements $\Sigma_{56}$ and $\Sigma_{65}$, respectively.
Finally, we have the complete scalar potential,
\begin{equation}\label{pEscalar}
 V=V_q+V_s+V_{B_{\mu}}.
\end{equation}

\noindent We need a potential for the Higgs doublets, so we minimize Eq. (\ref{pEscalar}) concerning $\sigma$ and substitute the result into Eq. (\ref{pEscalar}), obtaining the expression
\begin{equation}\label{potVH}
 V_H=\frac{1}{2}\left[m_1^2h_1^Th_1+m_2^2h_2^Th_2-2B_{\mu}h_1^Th_2+\lambda_0(h_1^Th_2)^2\right],
\end{equation}
where
\begin{equation}\label{potB}
 B_{\mu}=2\frac{\lambda_{56}m_{65}^2+\lambda_{65}m_{56}^2}{\lambda_{56}+\lambda_{65}},
\end{equation}
this expression can also be written in a more phenomenological form~\cite{kalyniak2015constraining}
\begin{equation}
 B_{\mu}=\frac{1}{2}(m_{A^0}^2+v^2\lambda_0)\sin(2\beta).
\end{equation}

\noindent The potential (\ref{potVH}) has a minimum when $m_1m_2 > 0$, on the other hand, EWSB requires $B_{\mu} > m_1m_2$ and $\lambda_0$ is the quartic coupling of the Higgs given by
\begin{equation}
 \lambda_0=2\frac{m^2_{h^0}}{v^2}\left(\frac{m^2_{h^0}-m^2_{A^0}}{m^2_{h^0}-m_{A^0}^2\sin^2(2\beta)}\right)
\end{equation}
The term $B_{\mu}$ disappears if $\lambda_{56}=0$ or $\lambda_{65}=0$ or both are zero in Eq. (\ref{potB}).
After EWSB, Higgs doublets acquires VEV's given by
\begin{equation}\label{aches}
 \langle h_1\rangle=v_1,\hspace{0.3cm}\langle h_2\rangle=v_2.
\end{equation}

\noindent The two terms in (\ref{aches}) must minimize Eq. (\ref{potVH}), resulting in the following relationships
\begin{eqnarray}
 v_1^2=\frac{1}{\lambda_0}\frac{m_2}{m_1}(B_{\mu}-m_1m_2),\\
 v_2^2=\frac{1}{\lambda_0}\frac{m_1}{m_2}(B_{\mu}-m_1m_2),
\end{eqnarray}
and it is defined the $\beta$ angle between $v_1$ and $v_2$~\cite{schmaltz2010bestest}, such that,
\begin{equation}
\tan\beta=\frac{\langle h_{11}\rangle}{\langle h_{21}\rangle}=\frac{v_1}{v_2}=\frac{m_2}{m_1},
\end{equation}
in this way, we have
\begin{eqnarray}
v^2&=&v_1^2+v_2^2\\\nonumber
&=&\frac{1}{\lambda_0}\left(\frac{m_1^2+m_2^2}{m_1m_2}\right)(B_{\mu}-m_1m_2)\\\nonumber
&\simeq&(246\;GeV)^2.
\end{eqnarray}
After the EWSB, the scalar sector~\citep{schmaltz2010bestest,phdthesis} produces massive states of $h^0$ (SM Higgs), $A^0$, $H^{\pm}$ and $H^0$, with masses
\begin{eqnarray}
\label{masa-esc1}
&&m^2_{G^0}=m^2_{G^{\pm}}=0,\\
\label{masa-esc2}
&&m^2_{A^0}=m^2_{H^{\pm}}=m^2_1+m^2_2,\label{masa-A0}\\
\label{masa-esc3}
&&m^2_{h^0,H^0}=\frac{B_{\mu}}{\sin2\beta}\label{masa-H0}\\\nonumber
\mp&&\sqrt{\frac{B^2_{\mu}}{\sin^22\beta}-2\lambda_0\beta_{\mu} v^2\sin2\beta+\lambda_0^2v^4\sin^22\beta},
\end{eqnarray}
where $G^0$ and $G^{\pm}$ are Goldstone bosons who are eaten to give masses to $W^{\pm},Z$ bosons of the SM.


\subsection{Gauge boson sector}
The gauge kinetic terms are given by the Lagrangian~\citep{schmaltz2010bestest,phdthesis}
\begin{equation}\label{lag-norma}
\mathcal{L}=\frac{f^2}{8}Tr\left(D_{\mu}\Sigma^{\dag}D^{\mu}\Sigma\right)+\frac{F^2}{4}Tr\left(D_{\mu}\Delta^{\dag}D^{\mu}\Delta\right),
\end{equation}
where $D_{\mu}\Sigma$ and $D_{\mu}\Delta$ are covariant derivatives.

The BLHM generates both the masses of the heavy gauge bosons and those of the SM bosons~\citep{schmaltz2010bestest,phdthesis},
\begin{eqnarray}\label{masas-boson}
m_{\gamma}^2&=&0,\\
m_{Z}^2&=&\frac{1}{4}v^2(g^2+g'\,^2)-(g^2+g'\,^2)\\\nonumber
&\times&\left(2+\frac{3f^2}{f^2+F^2}(s_g^2-c_g^2)\right)\frac{v^4}{48f^2},\\
m_W^2&=&\frac{1}{4}g^2v^2\\\nonumber
&-&g^2\Bigg(2+\frac{3f^2}{f^2+F^2}(s_g^2-c_g^2)\Bigg)\frac{v^4}{48f^2},\\
m_{Z'}^2&=&\frac{1}{4}(g_A^2+g_B^2)(f^2+F^2)-\frac{1}{4}g^2v^2\\\nonumber
&+&\Bigg(2g^2+\frac{3f^2}{f^2+F^2}\\
&\times&(g^2+g'\,^2)(s_g^2-c_g^2)\Bigg)\frac{v^4}{48f^2},\\
m_{W'}^2&=&\frac{1}{4}(g_A^2+g_B^2)(f^2+F^2)-M_W^2,
\end{eqnarray}
where $g'$ is the coupling of $U(1)_Y$, and $g$ of the $SU(2)_L$ they are related with the $SU(2)_A\times SU(2)_B$ couplings $g_A$ and $g_B$ in the way
\begin{eqnarray}
g=\frac{g_Ag_B}{\sqrt{g_A^2+g_B^2}},\\
s_g=\sin\theta_g=\frac{g_A}{\sqrt{g_A^2+g_B^2}},\\
c_g=\cos\theta_g=\frac{g_B}{\sqrt{g_A^2+g_B^2}},
\end{eqnarray}
here, $\theta_g$ is the mixing angle such that $g_A=g_B$ implies $\tan\theta_g=1$.


\subsection{Fermion sector}

The fermion sector of the BLHM is ruled by the Lagrangian~\cite{schmaltz2010bestest}
\begin{eqnarray}
\label{lag-yuk}
\mathcal{L}_t&=&y_1fQ^TS\,\Sigma\, SU^c+y_2fQ_a^{\prime T}\Sigma\,U^c\\\nonumber
&+&y_3fQ^T\Sigma\, U_5^{\prime c}+y_bfq_3^T(-2iT_ R^3\Sigma)U_b^c+\textrm{h.c.},
\end{eqnarray}
where $(Q,Q')$ and $(U,U')$ are multiplets of $SO(6)_A$ and $SO(6)_B$, respectively. $S$ is an operator of symmetry, $(y_1,y_2,y_3)$ are Yukawa couplings and $(q_3,U_b^c)$ in the last term, Eq. (\ref{lag-yuk}), contains information about bottom quark.
The BLHM is focused on the quark sector and proposes six heavy partner quarks: $T$, $T^5$, $T^6$, $T^{2/3}$, $T^{5/3}$ and $B$ with masses~\cite{schmaltz2010bestest}
\begin{eqnarray}
\label{masa-T}
m_{T}^2&=&(y_1^2+y_2^2)f^2\\\nonumber
&+&\frac{9v_1^2y_1^2y_2^2y_3^2}{(y_1^2+y_2^2)(y_2^2-y_3^2)},\\
\label{masa-T5}
m_{T^{5}}^2&=&(y_1^2+y_3^2)f^2\\\nonumber
&-&\frac{9v_1^2y_1^2y_2^2y_3^2}{(y_1^2+y_3^2)(y_2^2-y_3^2)},\\
\label{masa-T6}
m_{T^{6}}^2&=&m_{T^{2/3}}^2=m_{T^{5/3}}^2=y_1^2f^2,\\
\label{masa-B}
m_{B}^2&=&y_B^2f^2=(y_1^2+y_2^2)f^2,
\end{eqnarray}

\noindent where the Yukawa couplings in the quark sector Lagrangian~\cite{schmaltz2010bestest} must satisfy $0<y_i<1$. The masses of $t$ and $b$ are generated too by Yukawa couplings $y_t$ and $y_b$~\cite{phdthesis}
\begin{eqnarray}
m^2_t&=&y_t^2v_1^2,\label{masa-top}\\
m^2_b&=&y_b^2v_1^2-\frac{2y^2_b}{3\sin^2\beta}\frac{v^4_1}{f^2}.
\end{eqnarray}

The coupling $y_t$,
\begin{equation}\label{acople-yt}
 y_t^2=\frac{9y_1^2y_2^2y_3^2}{(y_1^2+y_2^2)(y_1^2+y_3^2)},
\end{equation}
is part of the measure of fine-tuning in the BLHM, $\Psi$, defined by~\cite{phdthesis}
\begin{equation}
\label{ajuste-fino}
\Psi=\frac{27f^2}{8\pi^2v^2\lambda_0\cos^2\beta}\frac{|y_1|^2|y_2|^2|y_3|^2}{|y_2|^2-|y_3|^2}\log\frac{|y_1|^2+|y_2|^2}{|y_1|^2+|y_3|^2}.
\end{equation}

%
%

\subsection{Flavor mixing in the BLHM}
The BLHM lacks fundamental FCNCs, similar to the SM. In the original development of the BLHM, the authors~\cite{schmaltz2010bestest} did not introduce interactions between the heavy quarks and the two light generations of SM quarks, which prevents the construction of extended CKM matrices. Therefore, we present the necessary terms with these interactions.
By adding the following terms to the Lagrangian (\ref{lag-yuk})
\begin{equation}\label{lag-ext1}
y_Bfq_1(-2iT_R^2\Sigma)d_B^c,\hspace{0.2cm}y_Bfq_2(-2iT_R^2\Sigma)d_B^c,
\end{equation}

\noindent where $y_B^2=y_1^2+y_2^2$ is the Yukawa coupling of heavy $B$ quark, $q_1$ and $q_2$ are multiplets of light SM quarks
\begin{eqnarray}\label{q1q2}
q_1^T&=&\frac{1}{\sqrt{2}}(-u,iu,d,id,0,0),\\\nonumber
q_2^T&=&\frac{1}{\sqrt{2}}(-c,ic,s,is,0,0),
\end{eqnarray}
and $d_B^c$ is the new multiplet that we have introduced
\begin{equation}
d_B^{c T}=(0,0,0,0,B,0).
\end{equation}
This allows us the mixing between scalars fields $(\phi^{\pm},\eta^{\pm},H^{\pm})$ and the $B$ with the $(u,c,d,s)$ quarks, which notably increase the phenomenology with the BLHM. The Lagrangian (\ref{lag-yuk}) keeps its gauge invariance, and the new terms do not mix with heavy partners of the top quark.
To obtain the interactions of the quarks $(u, c, d, s)$ with the $(W^{\pm}, W^{\prime\pm})$ bosons and the $B$ quark, we introduce the terms
\begin{equation}
Q_3^T=\frac{1}{\sqrt{2}}(0,0,B,iB,0,0),
\end{equation}
and
\begin{equation}
 q_i^{\prime T}=(0,0,0,0,q_i^c,0),
\end{equation}
where $q_i^c$ represents the quarks $(u, c, d, s)$, in the part of the Lagrangian for gauge-fermion interactions that involve the fields $W^{\pm}$ and $W^{\prime\pm}$~\cite{phdthesis}
\begin{eqnarray}\label{lag-corrientes}
\mathcal{L}&=&\sum_{i=1}^2 i\bar{\sigma}_{\mu}q^{\dag}_iD^{\mu}q_i+i\bar{\sigma}_{\mu}Q^{\dag}D^{\mu}Q\\\nonumber
&+& i\bar{\sigma}_{\mu}Q^{\prime\dag}D^{\mu}Q^{\prime}+i\bar{\sigma}_{\mu}U^{c\dag}D^{\mu}U^c,
\end{eqnarray}
such that we have the extended Lagrangian
\begin{eqnarray}\label{lag-corr-ex}
\mathcal{L}&=&\sum_{i=1}^2 i\bar{\sigma}_{\mu}q^{\dag}_iD^{\mu}q_i+i\bar{\sigma}_{\mu}Q^{\dag}D^{\mu}Q\\\nonumber
&+& i\bar{\sigma}_{\mu}Q^{\prime\dag}D^{\mu}Q^{\prime}+i\bar{\sigma}_{\mu}U^{c\dag}D^{\mu}U^c\\\nonumber
&+&\sum_{i=1}^2 i\bar{\sigma}_{\mu}Q_3^{\dag}D^{\mu}q_i+\sum_{i=1}^4 i\bar{\sigma}_{\mu}q^{\prime\dag}_{i}D^{\mu}U^c,
\end{eqnarray}
where $(q_1,q_2)$ are the same that Eqs. (\ref{q1q2}), $\bar{\sigma}^{\mu}=-\sigma^{\mu}$ are the Pauli matrix and $D_{\mu}$ contains information about $(W^{\pm},W^{\prime\pm})$. The extended Lagrangian (\ref{lag-corr-ex}) is gauge invariant and does not produce a mix with the heavy partners of the top quark. We can write the Lagrangian (\ref{lag-corr-ex}) in Dirac notation as follows:
\begin{eqnarray}\label{lag-ext-dirac}
 \mathcal{L}&=&i\sum_{i=1}^2\bar{\Psi}_{qi}\gamma_{\mu}P_LD^{\mu}\Psi_{qi}+i\bar{\Psi}_{Q}\gamma_{\mu}P_LD^{\mu}\Psi_{Q}\\\nonumber
 &+&i\bar{\Psi}_{Q'}\gamma_{\mu}P_LD^{\mu}\Psi_{Q'}+i\bar{\Psi}_{U^c}\gamma_{\mu}P_R\Psi_{U^c}\\\nonumber
 &+&i\sum_{i=1}^2\bar{\Psi}_{Q_3}\gamma_{\mu}P_LD^{\mu}\Psi_{qi}+i\sum_{i=1}^4\bar{\Psi}_{q'i}\gamma_{\mu}P_RD^{\mu}\Psi_{U^c}.
\end{eqnarray}
From Lagrangian (\ref{lag-ext-dirac}), any of its covariant derivatives in the mass eigenstate basis takes the form~\cite{phdthesis}:
\begin{eqnarray}
 D^{\mu}&=&\partial^{\mu}+i\sum_{a=1}^2g_AA^{a\mu}_1T^a_L\\\nonumber
 &=&\partial^{\mu}\\\nonumber
 &+&\frac{ig_A}{\sqrt{2}}\left[\rho_{11}(W^{+\mu}+W^{-\mu})+\delta_{11}(W^{\prime+\mu}+W^{\prime-\mu})\right]\\\nonumber
 &+&\frac{ig_A}{\sqrt{2}}\left[\rho_{12}(W^{+\mu}-W^{-\mu})+\delta_{12}(W^{\prime+\mu}-W^{\prime-\mu})\right],
\end{eqnarray}
where $\rho_{11}$ and $\delta_{11}$ contain model constants and dependence of $\mathcal{O}(v^2/(f^2+F^2))$. To demonstrate the generation of the elements of the extended CKM matrix $V_{Hu}$ in the charged currents of the BLHM, we choose the fifth term of Lagrangian (\ref{lag-ext-dirac}), and we consider the $W^{\prime-}\bar{B}q_i$ interaction:
\begin{equation}
 \frac{ig_A}{\sqrt{2}}\bar{\Psi}_BV_{Bu}\gamma_{\mu}P_LW^{\prime-}\Psi_{qi},
\end{equation}
where
\begin{equation}
 V_{Bu}=\delta_{11}-\delta_{12}\approx k\left(\frac{v^2}{f^2+F^2}\right),
\end{equation}
where $k$ depends on the model constants so that the values of the extended CKM matrix can constrain it.

%
%

\section{Parameter space of the BLHM}
\label{parametros}
The initial parameter spaces for the BLHM were based on the decays of heavy quarks into final states in the SM, considering an interdependent parametrization of the Yukawa couplings $(y_1, y_2, y_3)$~\cite{phdthesis,Martin:2012kqb}. The authors in \cite{Godfrey:2012tf} also use the same parameter space but only impose constraints on the heavy quarks according to the experimental limits. In \cite{kalyniak2015constraining}, a similar parameter space is proposed for the BLHM with the value $h^0=125$ GeV, allowing $A^0$ and $H^0$ to vary in all the production modes and decay channels investigated at ATLAS and CMS at that time. This rules out degeneracy between those fields.
In the articles \cite{Aranda:2021kza,Cruz-Albaro:2022kty,Cruz-Albaro:2022lks,Cruz-Albaro:2023pah}, a parameter space without degeneracy is constructed for the neutral Higgs states. The Yukawa couplings are not reparametrized, but the same values for the model constants are maintained as in the earlier articles. This agrees with the current experimental limits on heavy quarks and bosons.
The BLHM proposes five heavy partners of the top quark and only one heavy partner of the bottom quark. Experimental studies have excluded the masses of these particles below the limits shown in Table \ref{masas-quarks}.

\begin{table}[H]
\caption{Exotic quark mass searches in ATLAS~\cite{rappoccio2019experimental}.}
\centering
\label{masas-quarks}
\begin{tabular}{c c c c c c c c}
\toprule
Quark&  &Mass [TeV]&  \\
\midrule
$T$ &   & $1.37$ &  \\

$B$ & & $1.34$  & \\

$T^{5/3}$ & & $1.64$  &  \\
\bottomrule
\end{tabular}
\end{table}
In this work, we vary the angle of the ratio of the vacuum expectation values of the two Higgs doublets $(\tan\beta)$ to generate all non-free parameters while respecting the experimental constraints.

We begin by restricting the Yukawa couplings in the interval $(0 < y_i < 1)$~\cite{schmaltz2010bestest} to satisfy Eq. (\ref{masa-top}) and Eq. (\ref{acople-yt}), such that $m_t = 172.13$~\cite{CMS:2021jnp}. We simultaneously solve the Eqs. (\ref{masa-A0}) and (\ref{masa-H0}), such that $\lambda_0$, $m_{H^0}$, $m_{A^0}$, and $\beta$ satisfy $m_{h^0} = 125.46$~\cite{CMS:2020xrn}, $\lambda < 4\pi$, $\tan\beta \gtrsim 1$, and ~\cite{kalyniak2015constraining}

\begin{eqnarray}\label{beta}
\left(\tan\beta\right)^2&<&-1\\\nonumber
&+&\dfrac{2+2\left(1-\dfrac{m_{h^0}^2}{m_{A^0}^2}\right)^{\frac{1}{2}}\left(1-\dfrac{m_{h^0}^2}{4\pi v^2}\right)^{\frac{1}{2}}}{\dfrac{m_{h^0}^2}{m_{A^0}^2}\left(1+\dfrac{m_{A^0}^2-m_{h^0}^2}{4\pi v^2}\right)}.
\end{eqnarray}
The BLHM has the parameter $m_4$ as a free parameter, such that $m_{\phi}$ and $m_{\eta}$ are also free, considering that their one-loop corrections are very small~\cite{schmaltz2010bestest}. In this way, we choose $m_4 = 800$ GeV to ensure that the parameter space more comfortably satisfies the experimental masses in Table \ref{masas-quarks}.
The BLHM also consider the mixing angle $\alpha$ between $h^0$ and $H^0$,
\begin{eqnarray}
\label{ang-alfa}
&&\tan\alpha=\frac{1}{B_{\mu}-\lambda_0 v^2\sin2\beta}\Bigg(B_{\mu}\cot2\beta\\\nonumber
&+&\sqrt{\frac{B_{\mu}^2}{\sin^22\beta}-2\lambda_0 B_{\mu}v^2\sin2\beta+\lambda_0^2 v^4\sin^22\beta}\Bigg),
\end{eqnarray}
in such a way that the condition $\sin(\beta-\alpha)\approx1$ is satisfied.
Table \ref{parametros-BLH} presents the initial parameters with $y_1=0.7$ and $y_2=0.9$ held constant.

\begin{table}[H]
\caption{BLHM parameters.}
\centering
\label{parametros-BLH}
\begin{tabular}{c c c c c c c c}
\toprule
&Parameter & Interval& Units  \\
\midrule
& $\beta$ &   $(1.35,1.49)$ & rad \\

& $\alpha$ &   $(-0.42,-0.26)$ & rad \\

& $\lambda_0$ &  $(1,4\pi)$ & - \\

& $y_3$ &  $(0.32,0.33)$ & - \\

& $m_{A^0}$ &  $(0.30,1.69)$ & TeV\\

& $m_{H^0}$ &  $(0.91,1.90)$ & TeV \\

& $m_{\phi^0}$ &  $0.80$ & TeV\\

& $m_{\eta^0}$ &  $0.80$ & TeV\\
\bottomrule
\end{tabular}
\end{table}
The range of values for the angle $\beta$ was chosen within that interval because below it, tiny masses were generated for $A^0$ and $H^0$, and above it, these masses increased in value.

With the calculated Yukawa couplings and satisfying all the conditions that led us to the results in Table \ref{parametros-BLH}, we present the masses of the heavy quarks in Table \ref{masas-Quarks} according to equations (\ref{masa-T})-(\ref{masa-B}).

\begin{table}[H]
\caption{Masses of the heavy quarks in the BLHM.}
\centering
\label{masas-Quarks}
\begin{tabular}{c c c c c c c c}
\toprule
Quark& & Mass $(1<f<3)$ [TeV]&  \\
\midrule
$T$ &  &  $(1.15,3.42)$ &   \\

$T^5$ & & $(0.74,2.31)$  & \\

$T^6,T^{2/3},T^{5/3}$ & & $(0.70,2.10)$  & \\

$B$ & & $(1.14,3.42)$  & \\
\bottomrule
\end{tabular}
\end{table}

The most recent reports \cite{ATLAS:2023pja} with the same dataset from the period 2015-2018 from the ATLAS detector for the $T\to Ht$ and $T\to Zt$ processes exclude masses below 1.6 and 2.3 TeV for the production of the singlet $T$, according to the chosen parameters, and 1.7 and 1 TeV for the production of the doublet $T$ under the same considerations.
In \cite{ATLAS:2022tla}, masses of the $B$ quark below $1.33$ TeV are excluded according to the $B\to Wt$ process. In both cases, the masses of the top and bottom quark partners in the BLHM are within limits.

Interesting experimental results and simulations have been reported regarding the neutral bosons $H^0$ and $A^0$. In \cite{ATLAS:2015kpj}, the process $A^0\to Zh^0$ is analyzed at ATLAS, excluding the mass of $A^0$ in the range $(0.2,1)$ TeV for all types of Two-Higgs Doublet Model (2HDM). The study \cite{CMS:2019ogx} at CMS also excludes the mass of the $A^0$ boson below 1 TeV. In \cite{hashemi2022parameter}, the authors demonstrate that the alignment limit, $\cos(\beta-\alpha)=0$, in all 2HDM has not been excluded by the LHC, allowing for the possibility of proposing new parameter spaces to be analyzed with the results obtained in future colliders.
In \cite{hashemi2019search}, the type-I 2HDM are studied by simulating the $e^{-}e^{+}\to A^0H^0\to ZH^0H^0\to jjb\bar{b}b\bar{b}$ process for the SiD detector at the International Linear Collider (ILC) with an integrated luminosity of $500$ fb$^{-1}$, yielding the ranges $200<m_{A^0}<250$ GeV and $150<m_{H^0}<250$ GeV. The experimental evidence and simulations in the field of neutral scalar bosons allow us to use the ranges found for the masses of $A^0$ and $H^0$. The neutral scalar bosons $\phi^0$ and $\eta^0$ were introduced in the BLHM, considering they could have masses below $100$ GeV \cite{schmaltz2010bestest}. However, this type of light scalar mediator is more commonly found in models of dark matter \cite{ATLAS:2022ygn}. In the BLHM, the mass range $(30,800)$ GeV works well for $\phi^0$ and $\eta^0$, so we decided to take $800$ GeV as it increases the mass of $\phi^{\pm}$ and of $\eta^{\pm}$.

We have used $f\in[1,3]$ TeV and $F=5$ TeV as it is sufficient for the current experimental limits, as seen in Tables \ref{parametros-BLH} and \ref{masas-Quarks}. However, increasing those values to cover later results is possible as long as $f<F$, as the model requires. $F$ and $f$ contribute significantly to the masses of the vector bosons $W^{\prime\pm}$ and $Z^{\prime}$, as well as to the mass of the scalar bosons $(\phi^{\pm},\eta^{\pm})$. Using the calculated parameters, we display the masses of the vector and scalar bosons in Table \ref{masas-escalares}.

\begin{table}[H]
\caption{Masses of the heavy bosons in the BLHM.}
\centering
\label{masas-escalares}
\begin{tabular}{c c c c c c c c}
\toprule
Boson& & Mass $(1<f<3,\,F=5)$ [TeV]&  \\
\midrule
$m_{W'}$ &  &  $(3.32,3.80)$ &   \\

$m_{Z'}$ & & $(3.32,3.80)$  & \\

$m_{H^{\pm}}$ & & $(0.30,1.69)$  & \\

$m_{\phi^{\pm}}$ & & $(1.58,1.70)$  & \\

$m_{\eta^{\pm}}$ & & $(0.80,0.97)$  & \\
\bottomrule
\end{tabular}
\end{table}
Where the mass of the bosons $\phi^{\pm}$ and $\eta^{\pm}$ includes one-loop corrections~\cite{Martin:2012kqb},
\begin{eqnarray}
 m^2_{\phi^{\pm}}&=&\frac{16}{3}F^2\frac{3g_A^2g_B^2}{32\pi^2}log\left(\frac{\Lambda^2}{m_{W'}^2}\right)\\\nonumber
 &+&m_4^2\frac{f^4+f^2F^2+F^4}{F^2(f^2+F^2)}
\end{eqnarray}
and
\begin{equation}
 m^2_{\eta^{\pm}}=m_4^2+\frac{3f^2g^{\prime2}}{64\pi^2}\frac{\Lambda^2}{F^2}.
\end{equation}

\noindent In the $e\nu$ channel, the ATLAS and CMS collaborations set $m_{W'}$ in the $0.15-7$ TeV range and are based on $139$ fb$^{-1}$ at $\sqrt{s}= 13$ TeV~\cite{aad2019search,cms-pas}. These experimental ranges in the mass of the $W^{\prime\pm}$ boson keep the BLHM with good prospects because $W^{\prime\pm}$ can grow if $F$ does.
Searches for $Z'$ decays to $e^+e^-$ and $\mu^+\mu^-$ in ATLAS and CMS collaborations \cite{ATLAS:2019erb,CMS:2021ctt} set a lower limit of $4.9$ TeV for the mass of $Z'$. On the other hand, the $Z'\to\tau^+\tau^-$ decay imposes a limit $m_{Z'}>2.4$ TeV~\cite{ATLAS:2017eiz}. In the $Z'\to b\bar{b}$ decay, CMS \cite{cms-jul} found the range $1.8<m_{Z'}<8$ TeV, and ATLAS~\cite{ATLAS:2019fgd} found the range $1.3<m_{Z'}<5$ TeV. In all the mentioned cases, our results for the mass of $Z'$ can be appropriately accommodated.

The mass of the charged scalar boson $H^{\pm}$ also arises naturally from the variation of the $\beta$ angle, and the obtained range covers experimental studies in processes such as $H^{\pm}\to h^0W^{\pm}$ with results in the interval $(0.3, 0.7)$ TeV~\cite{cms2022search}.

\section{Phenomenology of flavor-changing top quark rare decays}
\label{sp}

\begin{figure}[H]
\centering
\includegraphics[width=8.5cm]{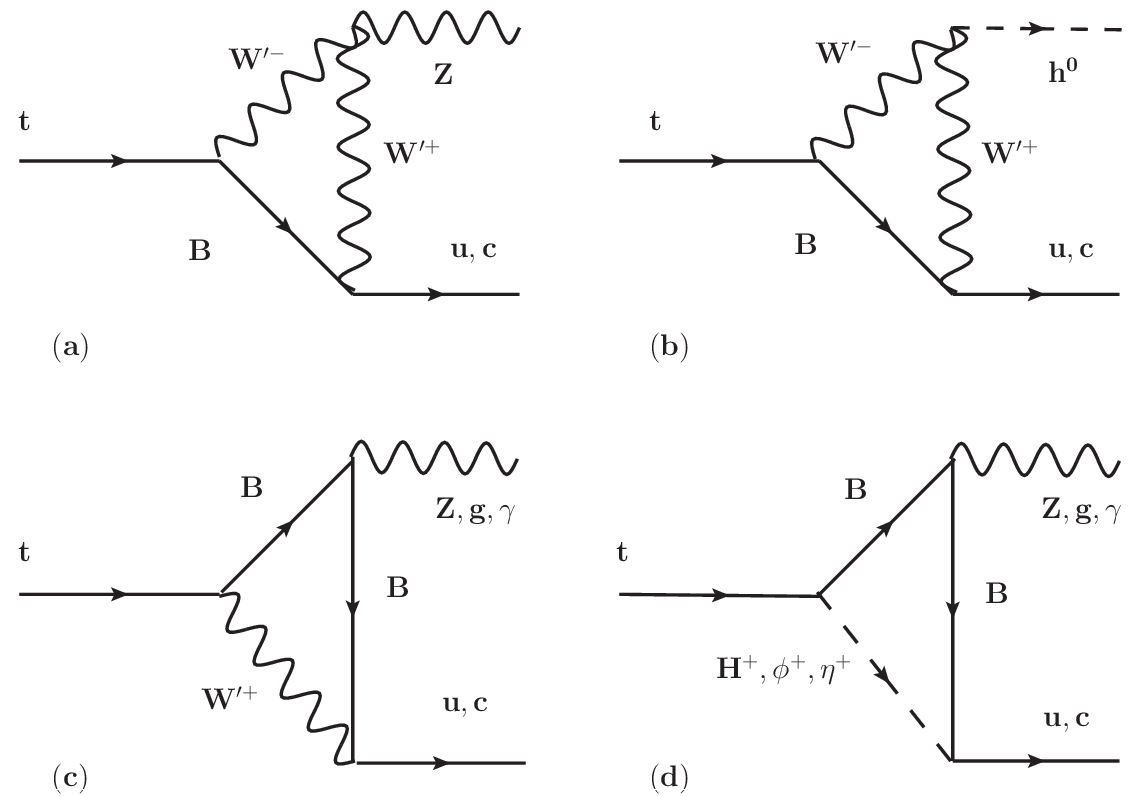}
\caption{Feynman diagrams for the flavor-changing top quark rare decays in the BLHM considered in this paper: $t\rightarrow qV$ and $t \rightarrow qh^o$ vertices, with $V= Z,\gamma, g$, and $q= u, c$.}
\label{diagramas-fcnc}
\end{figure}

\begin{figure}[H]
\centering
\includegraphics[width=8.5cm]{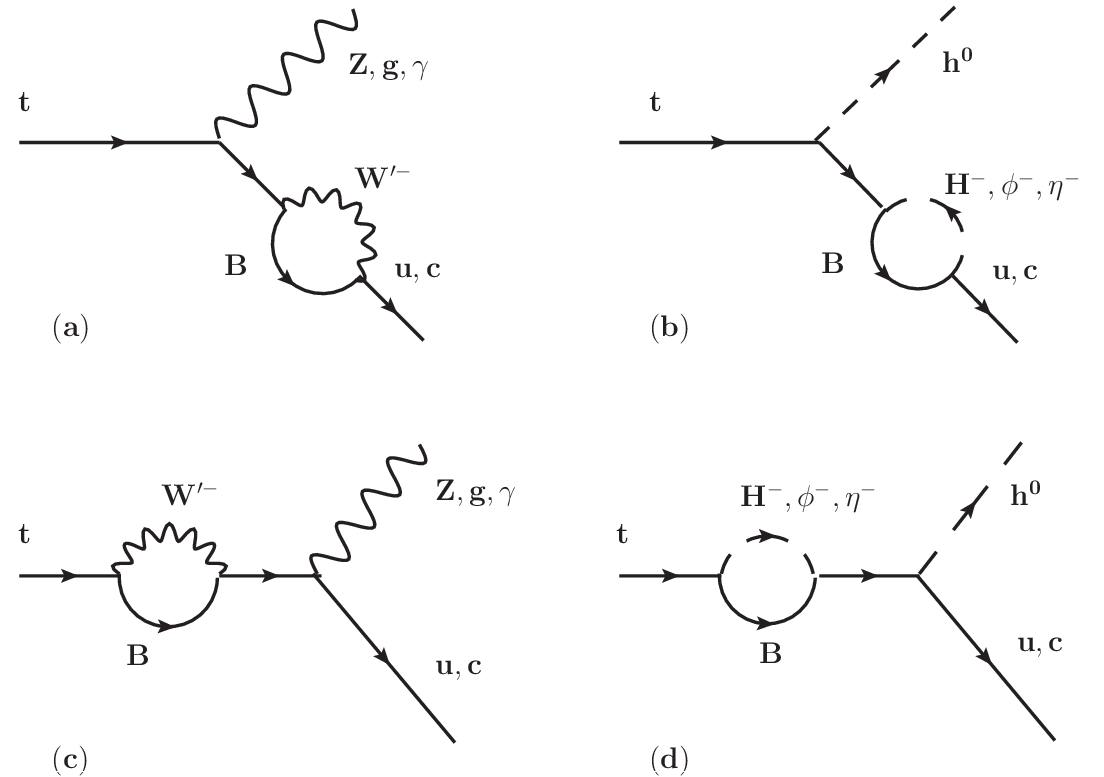}
\caption{Feynman diagrams for the flavor-changing top quark rare decays in the BLHM considered in this paper: $t\rightarrow qV$ and $t \rightarrow qh^o$ vertices, with $V= Z,\gamma, g$, and $q= u, c$.}
\label{diagramas-fcnc2}
\end{figure}

The valid diagrams for $t\rightarrow qV$ and $t\rightarrow qh^0$ where $q=(u,c)$ and $V=(Z,\gamma,g)$ are showed in Figs. (\ref{diagramas-fcnc}) and (\ref{diagramas-fcnc2}).  We have calculated fifty-two amplitudes for each matrix in Case II and Case III using the Mathematica packages FeynCalc~\cite{shtabovenko2020feyncalc} and Package X~\cite{patel2015package}. Each amplitude of the decay $t\rightarrow qV$ has the structure
\begin{equation}
\mathcal{M}^{\mu} = \bar{u}(p_j)\left(
F_1p_i^\mu\mathbf{1}+F_2p_i^\mu\gamma^5+F_3\gamma^\mu+F_4\gamma^\mu\gamma^5
\right)u(p_i).
\end{equation}

\noindent The form factors $(F_1, F_2, F_3, F_4)$ include the masses and momenta of the external and internal quarks and gauge bosons in the BLHM and the SM within the Passarino-Veltman scalar functions. Each amplitude exhibits a distinct structure in the case of $t\rightarrow qh^0$.

\begin{equation}
\mathcal{M}^{\mu} = \bar{u}(p_j)\left(f_1+f_2\gamma^5\right)u(p_i),
\end{equation}

\noindent $f_{1,2}$ contains terms from the BLHM and SM with Passarino-Veltman scalar functions.

\subsection{Cases for the CKM matrix in the BLHM}
\label{cases}

In various BSM, the mass eigenstates do not necessarily align with those of the SM, which introduces new contributions from both vector and scalar fields to processes involving FCNC. One advantage of these models is that they are not subject to suppression through the GIM mechanism \cite{glashow1970weak}. In the BLHM, radiative corrections must be included to quantify flavor violation, as it lacks it at the tree level, similar to the SM. This study uses the same method used in \cite{blanke2006particle,blanke2007rare,blanke2009fcnc} to construct the extended CKM matrices.
We need two CKM-like unitary matrices
\begin{equation}\label{ckm-like}
V_{Hu},\;V_{Hd}
\end{equation}
such that
\begin{equation}
V_{CKM}=V_{Hu}^{\dagger}V_{Hd}.
\end{equation}
Actually, we know $V_{CKM}$ therefore $V_{Hd}=V_{Hu}V_{CKM}$. The matrices in (\ref{ckm-like}) parameterize flavor violating interactions between SM fermions $(u,c)$ and bosons $(Z,\gamma,g,h^0)$, and BLHM fermion $B$ mediated by $(W^{\prime\pm},\phi^{\pm},\eta^{\pm},H^{\pm})$, in this paper. We can generalize the CKM extended matrix like the product of three rotations matrices~\citep{blanke2007another,blanke2007rare}
\begin{eqnarray}
V_{Hd}&&=\begin{pmatrix}
1&0&0\\
0&c_{23}^d&s_{23}e^{-i\delta_{23}^d}\\
0&-s_{23}^de^{i\delta_{23}^d} &c_{23}^d
\end{pmatrix}\\\nonumber
&&\times \begin{pmatrix}
c_{13}^d&0&s_{13}^de^{-i\delta_{13}^d}\\
0&1&0\\
-s_{13}e^{i\delta_{13}^d}&0&c_{13}^d
\end{pmatrix}\\\nonumber
&&\times \begin{pmatrix}
c_{12}^d&s_{12}^de^{-i\delta_{12}^d}&0\\
-s_{12}^de^{i\delta_{12}^d}&c_{12}^d&0\\
0&0&1
\end{pmatrix}.
\end{eqnarray}

\noindent Where the $c^d_{ij}$ and $s^d_{ij}$ are in terms of the angles $(\theta_{12},\theta_{23},\theta_{13})$ and the phases $(\delta_{12},\delta_{23},\delta_{13})$.
We choose the following cases:\par\par
\noindent\textbf{Case I.} $V_{Hu}=\mathbf{1}$, this implies $V_{Hd}=V_{CKM}$.\par

In this case, the condition $V_{Hu}=\mathbf{1}$ does not allow for the contribution of the extended matrix or the CKM matrix since there are no quarks or bosons from the SM within the calculated loops. Only the model constants and the masses of the charged scalar bosons will be present.

\noindent\textbf{Case II.} $V_{Hd}=\mathbf{1}$, this implies $V_{Hu}=V_{CKM}^{\dagger}$.\par

In this case, the contribution comes from the CKM matrix, in which we will have suppression due to smaller terms such as $V_{ub}$ and $V_{cb}$.

\noindent\textbf{Case III.}\label{caso3} $s_{23}^d=1/\sqrt{2}$, $s_{12}^d=s_{13}^d=0$, $\delta_{12}^d=\delta_{23}^d=\delta_{13}^d=0$.\par

In this case, we follow the parametrization in \cite{blanke2007rare,blanke2006particle}. Of all the matrix elements, $|V_{Bu}|=0.006$ is the only one suppressed. However, the branching ratios of the decays turned out to be much larger than their counterparts in the SM. We have yet to consider other scenarios for the extended CKM matrix because one of the objectives of this study was to broaden the flavor structure in the BLHM. Future studies on CP violation will consider other cases.

%
%

\subsection{Branching ratios for the reactions \texorpdfstring{$t \to qV$ and $t \to qh^0$}\\ in the BLHM.}

Using the parameter space detailed in Section \ref{sp} and the matrix $V_{Hu}$ from the second and third cases in Section \ref{cases}, we have calculated the branching ratios of the processes $t\to qV$ and $t\to qh^0$, where $q=u,c$ and $V=Z,\gamma, g$. Our results are shown in Tables \ref{branchings2} and \ref{branchings}.

\subsubsection{\textbf{Case II}}

In this section, we present the branching ratios calculated using the extended matrix $V_{Hu}=V_{CKM}^{\dag}$. The numerical results for the top quark decay branching ratios are listed in Table \ref{branchings2}, where we observe that the smallest ratios involve the up quark. The decays involving the charm quark are also small compared to those obtained through the matrix of Case III. Thus, it is evident that the contribution of the extended matrix and the model itself increase the branchings.

\begin{table}[H]
\centering
\caption{\textbf{Case II.} The branching ratios for the top-quark decay via flavor-changing neutral current couplings at the BLHM.}
\begin{tabular}{|c|c|c|}\cline{1-3}
\rule{0pt}{4ex} & $Br_{BLH}$ & $Br_{BLH}$ \\\cline{1-3}
\rule{0pt}{4ex} Decay & $f=1$ TeV & $f= 3$ TeV \\\cline{1-3}
\rule{0pt}{4ex} $t\rightarrow uZ$ & $5.8\times10^{-10}$ & $5.0\times10^{-11}$ \\\cline{1-3}
\rule{0pt}{4ex}$t\rightarrow u\gamma$ & $4.0\times10^{-11}$ & $3.6\times10^{-12}$\\\cline{1-3}

\rule{0pt}{4ex}$t\rightarrow ug$ & $6.7\times10^{-17}$ & $7.2\times10^{-20}$\\\cline{1-3}

\rule{0pt}{4ex}$t\rightarrow uh^0$ & $1.3\times10^{-12}$ & $9.8\times10^{-15}$\\\cline{1-3}

\rule{0pt}{4ex}$t\rightarrow cZ$ & $1.2\times10^{-8}$ & $1.0\times10^{-9}$\\\cline{1-3}

\rule{0pt}{4ex}$t\rightarrow c\gamma$ & $8.8\times10^{-10}$ & $7.9\times10^{-11}$\\\cline{1-3}

\rule{0pt}{4ex}$t\rightarrow cg$ & $1.4\times10^{-15}$ & $1.5\times10^{-18}$\\\cline{1-3}

\rule{0pt}{4ex}$t\rightarrow ch^0$ & $2.9\times10^{-11}$ & $2.1\times10^{-13}$\\\cline{1-3}
\end{tabular}
\label{branchings2}
\end{table}

\begin{figure}[H]
\centering
\includegraphics[width=7cm]{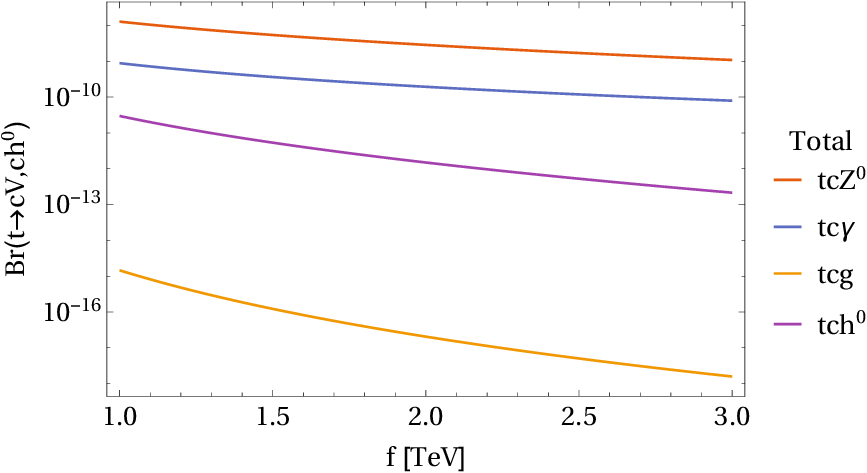}
\caption{\textbf{Case II.} Total contributions to $Br(t\rightarrow cZ, c\gamma, cg, ch^0)$ as a function of the scale of energy $f$, with $m_B=(1.14)f$ and $F=5$ TeV.}
\label{charm2}
\end{figure}

\begin{figure}[H]
\centering
\includegraphics[width=7cm]{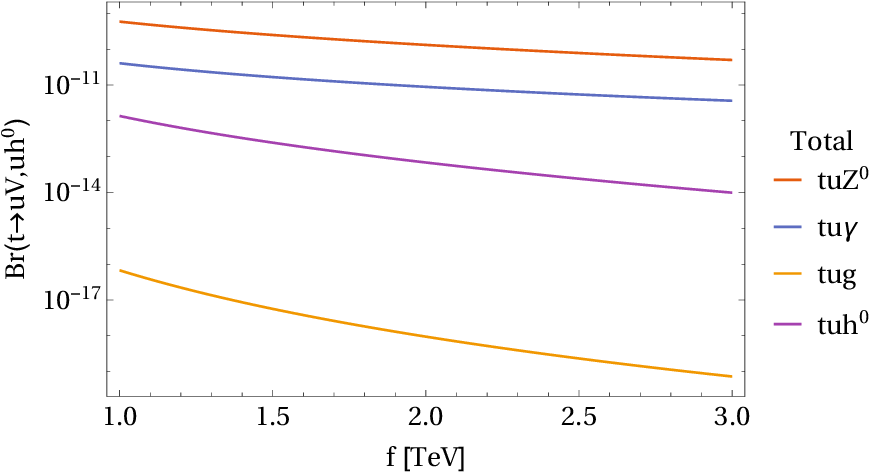}
\caption{\textbf{Case II.} Total contributions to $Br(t\rightarrow uZ, u\gamma, ug, uh^0)$ as a function of the scale of energy $f$, with $m_B=(1.14)f$ and $F=5$ TeV.}
\label{up2}
\end{figure}

 In Fig. \ref{charm2}, we plot $B(t\to cV,ch^0)$ against the breaking scale $f$, with lower intensities compared to the equivalent branchings in Fig. \ref{charm}. In Fig. \ref{up2}, we observe a different situation as the $B(t\to uV,uh^0)$ have similar magnitudes to their counterparts in Fig. \ref{up}, indicating that the form of the extended matrix for Case III practically did not increase the sensitivity in the up quark branching ratios.

This difference in contributions from the extended matrices will allow us to study more cases where obtaining larger magnitudes in these same decays relative to experimental limits is possible.

\subsubsection{\textbf{Case III}}

In this section, we present the branching ratios for the extended matrix of Case III, along with the corresponding graphs. We consider it necessary to showcase the individual contributions of the fields $(W^{\prime\pm},\phi^{\pm},\eta^{\pm},H^{\pm})$ in the BLHM to the branching of each field $(Z,\gamma,g,h^{0})$ and the quarks $u, c, B$ in Appendix \ref{contribuciones}.
We can observe that the scalar field $\phi^{\pm}$ was the most significant contribution to the branching ratios for both decays processes $t\to cV$ and $t\to ch^0$. This highlights the significance of the scalar field contributions in the BLHM in conjunction with the extended CKM matrix.

\begin{table}[H]
\centering
\caption{\textbf{Case III.} The branching ratios for the top-quark decay via flavor-changing neutral current couplings at the BLHM.}
\begin{tabular}{|c|c|c|}\cline{1-3}
\rule{0pt}{4ex} & $Br_{BLH}$ & $Br_{BLH}$ \\\cline{1-3}
\rule{0pt}{4ex} Decay & $f=1$ TeV & $f= 3$ TeV \\\cline{1-3}
\rule{0pt}{4ex}$t\rightarrow uZ$ & $3.5\times10^{-10}$ & $3.0\times10^{-11}$\\\cline{1-3}

\rule{0pt}{4ex}$t\rightarrow u\gamma$ & $2.5\times10^{-11}$ & $2.2\times10^{-12}$\\\cline{1-3}

\rule{0pt}{4ex}$t\rightarrow ug$ & $4.0\times10^{-17}$ & $4.2\times10^{-20}$\\\cline{1-3}

\rule{0pt}{4ex}$t\rightarrow uh^0$ & $8.2\times10^{-13}$ & $5.9\times10^{-15}$\\\cline{1-3}

\rule{0pt}{4ex}$t\rightarrow cZ$ & $3.7\times10^{-5}$ & $3.1\times10^{-6}$\\\cline{1-3}

\rule{0pt}{4ex}$t\rightarrow c\gamma$ & $2.6\times10^{-6}$ & $2.3\times10^{-8}$\\\cline{1-3}

\rule{0pt}{4ex}$t\rightarrow cg$ & $4.2\times10^{-13}$ & $4.5\times10^{-16}$\\\cline{1-3}

\rule{0pt}{4ex}$t\rightarrow ch^0$ & $8.5\times10^{-9}$ & $6.2\times10^{-11}$\\\cline{1-3}
\end{tabular}
\label{branchings}
\end{table}

The best sensitivity on the branching ratios might reach up to the order of magnitude of $\mathcal{O}(10^{-8}-10^{-5})$. As can be seen from Table \ref{branchings}, the decay channels $t \rightarrow cZ$, $t \rightarrow c\gamma$, and $t \rightarrow ch^0$ exhibit good sensitivities compared to the branching ratios reported in the literature~\cite{hong2007flavor,han2016revisiting}.
Experimentally, the data provided by ATLAS and CMS place the latest measurements of the branching ratios for $t\to qZ$ with a limit of $10^{-4}$ \cite{Aaboud2018}. In \cite{atlas2022search}, for the decay of the top quark to a photon accompanied by a charm quark or an up quark, they report $Br(t\to q\gamma)\sim10^{-5}$. For the process $t\to qg$, the branching ratio is around $10^{-4}$ \cite{ATLAS:2021amo}. In \cite{Tumasyan2022}, the estimation $Br(t\to qh^0)\sim10^{-3}$ is published.

\begin{figure}[H]
\centering
\includegraphics[width=7cm]{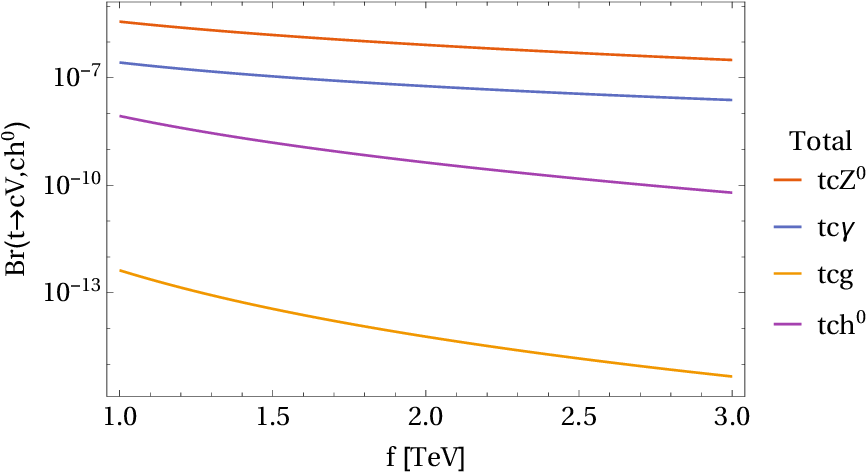}
\caption{\textbf{Case III.} Total contributions to $Br(t\rightarrow cZ, c\gamma, cg, ch^0)$ as a function of the scale of energy $f$, with $m_B=(1.14)f$ and $F=5$ TeV.}
\label{charm}
\end{figure}

Fig. \ref{charm} shows the branching ratio of the decays $t\to cV$ and $t\to ch^0$, where $V=(Z,\gamma,g)$, as a function of the breaking scale energy $f$ in the range $[1,3]$ TeV. Since the masses of the bosons $(W^{\prime\pm},\phi^{\pm},\eta^{\pm},H^{\pm})$ and the quark $B$ depend on the scale $f$, they also vary their contribution for each point of the branching ratio in the plot of Fig. \ref{charm}.
It is important to emphasize that for the parameters $(\beta,\alpha,\lambda_0,y_3)$, their lower limits given in Table \ref{parametros-BLH} were used. In general, we could calculate the branchings for any value of $\beta$ within the range $[1.35, 149]$, but even though the masses of $A^0$ and $H^0$ would be more significant, this would not significantly alter the scenario in the plot of Fig.~\ref{charm} since the masses of $W^{\prime\pm}$ and $Z'$ reach much higher values as a function of $f$ and $F$.

\begin{figure}[H]
\centering
\includegraphics[width=7cm]{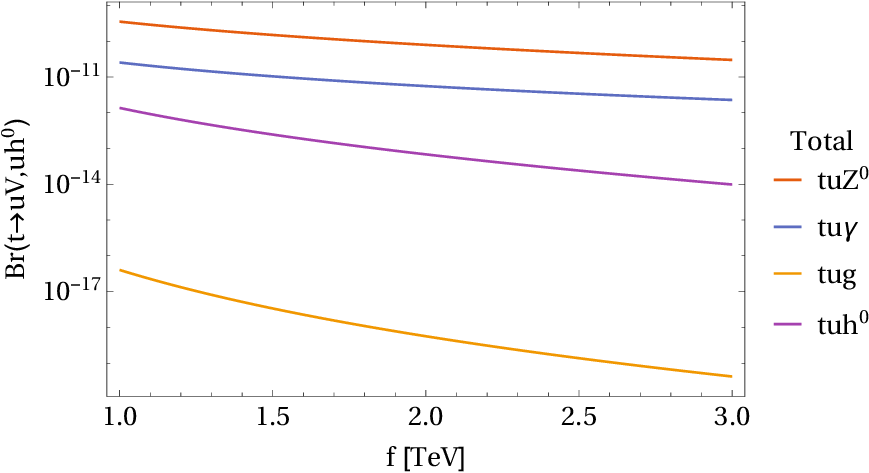}
\caption{\textbf{Case III.} Total contributions to $Br(t\rightarrow uZ, u\gamma, ug, uh^0)$ as a function of the scale of energy $f$, with $m_B=(1.14)f$ and $F=5$ TeV.}
\label{up}
\end{figure}

In the case of the decays $t\to uV$ and $t\to uh^0$, under the same parameters, the branching ratios were found to be more suppressed, as shown in Fig. \ref{up}. Given the records mentioned in the literature, it was expected that the BLHM would produce similar ranges for these decays. If we compare our branching ratios with those obtained in \cite{hong2007flavor,han2016revisiting}, in the LHT model, we can observe that their predictions for $Br(t\to cg)\sim10^{-2}$ and $Br(t\to c\gamma)\sim10^{-2}$ have already been surpassed in experimental searches \cite{ATLAS:2021amo,atlas2022search}.

\section{Conclusions}
\label{conclusiones}

In this paper, we calculate and study the effects of the one-loop contributions from the heavy gauge bosons $W^{\prime\pm}$, heavy quark bottom $B$, and the new scalars $(\phi^{\pm},\eta^{\pm},H^{\pm})$ predicted for the BLHM, as well as of the effect of the parameters from this model on the top quark rare decay processes $t\rightarrow qV$ and $t\rightarrow qh^0$.
Our results for $Br(t \to qV)$ and $Br(t \to qh^0)$ are summarized in Tables \ref{branchings}--\ref{branchings2} and illustrated in Figs. (\ref{charm2})-(\ref{Hexp}).
From these Tables and the Figures, it is evident that the most significant contributions to the branching ratios are observed in the channels $Br(t\rightarrow cZ)=3.7\times10^{-5}$, $Br(t\to c\gamma)=2.6\times10^{-6}$, and $Br(t\to ch^0)=8.5\times10^{-8}$, considering a symmetry-breaking scale of $f=[1,3]$ TeV in the BLHM.
The rare decay modes $t \to cZ$ and $t\to c\gamma$ investigated in this article can be measured with good sensitivity in the High Luminosity (HL) and High Energy (HE) era at the LHC~\cite{feng2023forward}, as well as in the Future Circular hadron-hadron Collider (FCC-hh). Future lepton colliders such as the Compact Linear Collider (CLIC)\cite{JOUR,CLICdp:2018cto} and the muon Collider with HL and HE\cite{AlAli_2022} also consider the physics of the top quark in their research programs.
In summary, this study analyzes the prospect of constraining the branching ratios $Br(t \to uZ, u\gamma, ug, uh^0)$ and $Br(t \to cZ, c\gamma, cg, ch^0)$ in the BLHM. Additionally, this study provides the first estimate for constraining these couplings in the BLHM. Given its theoretical and phenomenological characteristics, the BLHM is considered a promising candidate for future experiments in the medium-term.

\begin{acknowledgments}
A CONACYT Postdoctoral Fellowship has supported this work.
\end{acknowledgments}

\appendix

\section{Individual Contributions from \texorpdfstring{$W^{\prime\pm},\phi^{\pm},\eta^{\pm},H^{\pm}$}\\.}
\label{contribuciones}

In this appendix, we present the graphs corresponding to the individual contributions of the fields $(W^{\prime\pm},\phi^{\pm},\eta^{\pm},H^{\pm})$ in the BLHM to each field $(Z,\gamma,g,h^0)$ in the SM, where the quarks $(u, c)$ and $B$ are also involved.

\begin{figure}[H]
\centering
\includegraphics[width=8cm]{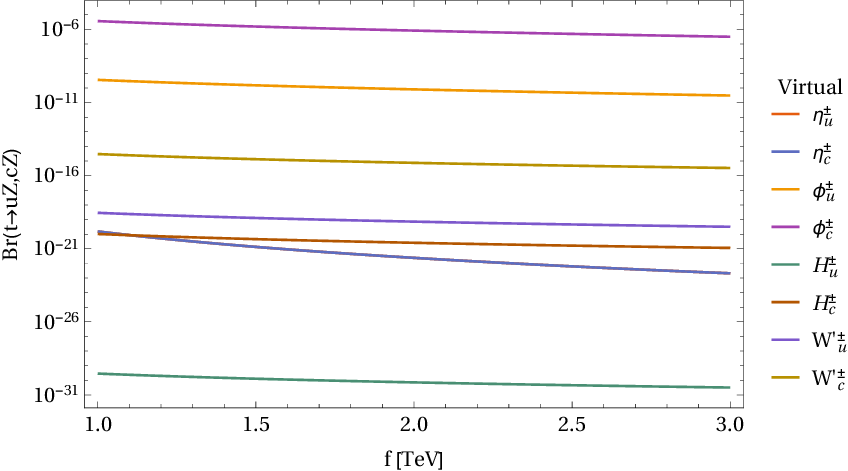}
\caption{Individual contributions from $W^{\prime\pm},\phi^{\pm},\eta^{\pm},H^{\pm}$ to $Z$.}
\label{Zexp}
\end{figure}

\begin{figure}[H]
\centering
\includegraphics[width=8cm]{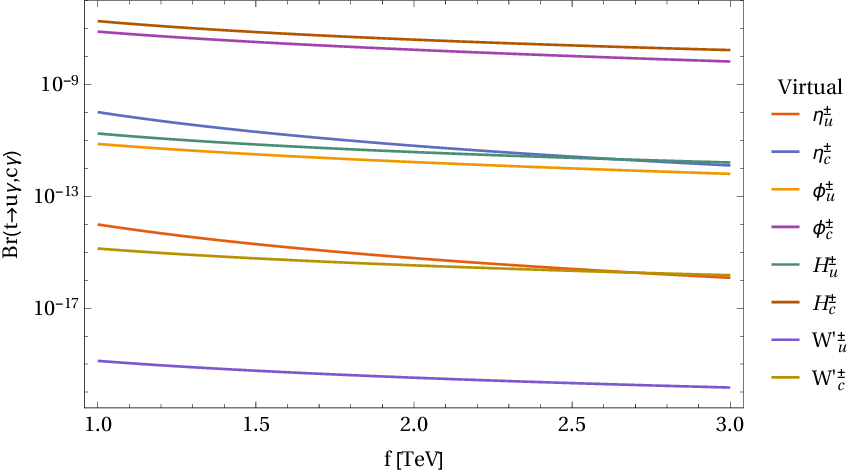}
\caption{Individual contributions from $W^{\prime\pm},\phi^{\pm},\eta^{\pm},H^{\pm}$ to $\gamma$.}
\label{Fexp}
\end{figure}

\begin{figure}[H]
\centering
\includegraphics[width=8cm]{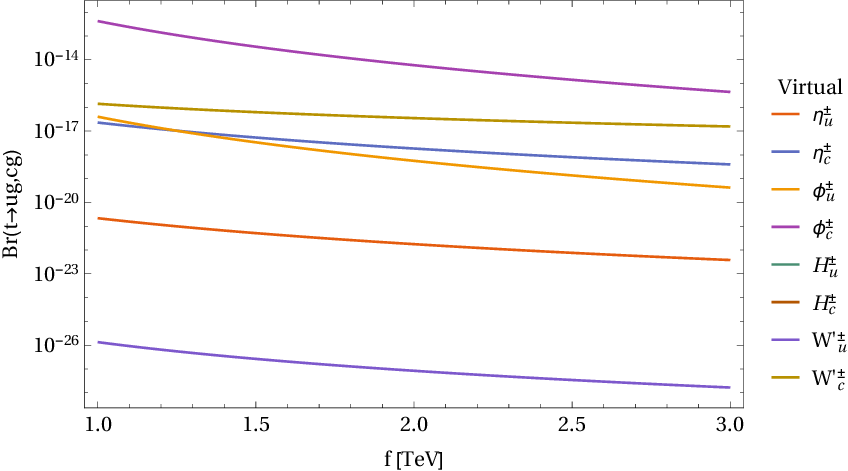}
\caption{Individual contributions from $W^{\prime\pm},\phi^{\pm},\eta^{\pm},H^{\pm}$ to $g$.}
\label{Gexp}
\end{figure}

\begin{figure}[H]
\centering
\includegraphics[width=8cm]{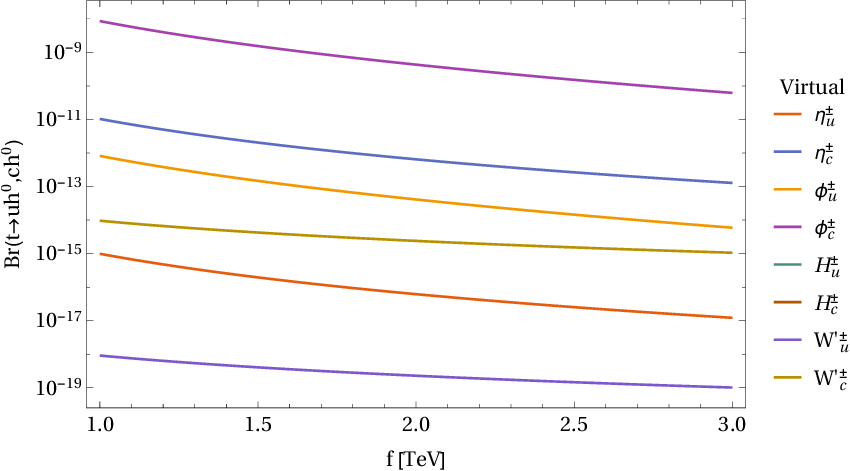}
\caption{Individual contributions from $W^{\prime\pm},\phi^{\pm},\eta^{\pm},H^{\pm}$ $h^0$.}
\label{Hexp}
\end{figure}

In Fig. \ref{Zexp}, the contributions of $\phi^{\pm}_c$ dominate. In Fig. \ref{Fexp}, $H_c^{\pm}$ and $\phi_c^{\pm}$ are the largest contributors. In Fig. \ref{Gexp}, $\phi_c^{\pm}$ is the dominant contributor. Finally, in Fig. \ref{Hexp}, $\phi_c^{\pm}$ has the highest contribution to the branching ratio.

\section{Feynman rules in the BLHM}
\label{feynrules}

In this appendix, we derive and present the Feynman rules for the BLHM necessary to calculate the flavor-changing top quark rare decays.

Tables \ref{feynrulesTabla}--\ref{feynrulesTabla4} summarize the Feynman rules for the 3-point interactions: fermion-fermion-scalar (FFS), fermion-fermion-gauge (FFV), gauge-gauge-gauge (VVV), and scalar-gauge-gauge (SVV) interactions.

\begin{table}[H]
\centering
\caption{Essential Feynman rules in the BLHM for studying the flavor-changing top quark rare decays are the 3-point interactions fermion-fermion-scalar (FFS), fermion-fermion-gauge (FFV), gauge-gauge-gauge (VVV), and scalar-gauge-gauge (SVV) interactions.}
\begin{tabular}{|c|c|}\cline{1-2}
\rule{0pt}{4ex} Vertex & Rule \\\cline{1-2}
\rule{0pt}{4ex} $W^{\prime -}\bar{B}t$ & $\dfrac{-igs_{\beta}v}{4\sqrt{2}f}A_y\gamma^{\mu}P_L(V_{Hu})$ \\\cline{1-2}
\rule{0pt}{4ex} $W^{\prime -}\bar{B}u$ & $\dfrac{igg_A}{2\sqrt{2}g_B}\gamma^{\mu}P_L(V_{Hu})$ \\\cline{1-2}
\rule{0pt}{4ex} $W^{\prime -}\bar{B}c$ & $\dfrac{igg_A}{2\sqrt{2}g_B}\gamma^{\mu}P_L(V_{Hu})$ \\\cline{1-2}
\rule{0pt}{4ex} $\eta^{-}\bar{B}t$ & $\dfrac{4im_W^2}{f^2g^2\sqrt{y_1^2+y_2^2}(y_1^2+y_3^2)}(Y_1P_L+Y_2P_R)(V_{Hu})$ \\\cline{1-2}
\rule{0pt}{4ex} $\eta^{-}\bar{B}u$ & $-\dfrac{im_Bs_{\beta}^2}{2f\sqrt{2}}P_L(V_{Hu})$ \\\cline{1-2}
\rule{0pt}{4ex} $\eta^{-}\bar{B}c$ & $-\dfrac{m_Bs_{\beta}^2}{2f\sqrt{2}}P_L(V_{Hu})$ \\\cline{1-2}
\rule{0pt}{4ex} $\phi^{-}\bar{B}t$ & $F_aF_b(X_1P_L+X_2P_R)(V_{Hu})$ \\\cline{1-2}
\rule{0pt}{4ex} $\phi^{-}\bar{B}u$ & $\dfrac{iFs^2_{\beta}\left[m_u+m_B+(m_u-m_B)\gamma^5\right]}{2f\sqrt{2}\sqrt{f^2+F^2}}(V_{Hu})$ \\\cline{1-2}
\rule{0pt}{4ex} $\phi^{-}\bar{B}c$ & $\dfrac{iFs^2_{\beta}\left[m_c+m_B+(m_c-m_B)\gamma^5\right]}{2f\sqrt{2}\sqrt{f^2+F^2}}(V_{Hu})$ \\\cline{1-2}
\rule{0pt}{4ex} $H^{-}\bar{B}t$ & $\dfrac{-3\sqrt{2}m_Wc_{\beta}s_{\beta}y_1y_2y_3^2}{fg\sqrt{y_1^2+y_2^2}(y_1^2+y_3^2)}
P_L(V_{Hu})$ \\\cline{1-2}
\rule{0pt}{4ex} $H^{-}\bar{B}u$ & $\dfrac{gm_{B}s_{2\beta}}{4\sqrt{2}m_W}P_L(V_{Hu})$ \\\cline{1-2}
\rule{0pt}{4ex} $H^{-}\bar{B}c$ & $\dfrac{gm_{B}s_{2\beta}}{4\sqrt{2}m_W}P_L(V_{Hu})$ \\\cline{1-2}
\rule{0pt}{4ex} $Z\bar{B}B$ & $-\dfrac{ig}{6c_W}(3c_W^2-7s^2_W)\gamma^{\mu}$ \\\cline{1-2}
\rule{0pt}{4ex} $\gamma\bar{B}B$ & $-\dfrac{1}{3}igs_{W}\gamma^{\mu}$\\\cline{1-2}
\rule{0pt}{4ex} $Z_{q,\mu}W'^{-}_{k,\alpha}W'^{+}_{p,\beta}$ & $igc_W\left[\delta_{\beta\mu}\Delta_1+\delta_{\alpha\mu}\Delta_2+\delta_{\alpha\beta}\Delta_3\right]$ \\\cline{1-2}
\rule{0pt}{4ex} $h^0W'^{-}W'^{+}$ & $2gm_{W}s_{\alpha+\beta}$\\\cline{1-2}
\end{tabular}
\label{feynrulesTabla}
\end{table}

\begin{table}[H]
\centering
\caption{Factors from Table \ref{feynrulesTabla}.}
\begin{tabular}{|c|c|}\cline{1-2}
\rule{0pt}{4ex} Factor & Expression \\\cline{1-2}
\rule{0pt}{4ex} $A_y$ & $y_3(2y_1^2-y_2^2)(y_1^2+y_2^2)^2(y_1^2+y_3^2)$ \\\cline{1-2}
\rule{0pt}{4ex} $Y_1$ & $B_1+B_2$ \\\cline{1-2}
\rule{0pt}{4ex} $Y_2$ & $B_3+B_4$ \\\cline{1-2}
\rule{0pt}{4ex} $F_a$ & $\dfrac{y_3(2y_1^2-y_2^2)}{(y_1^2+y_3^2)\sqrt{y_1^2+y_2^2}}$ \\\cline{1-2}
\rule{0pt}{4ex} $F_b$ & $\dfrac{Fs_{\beta}}{2f\sqrt{2(f^2+F^2)}}$ \\\cline{1-2}
\rule{0pt}{4ex} $X_1$ & $A_1+A_3+A_6$ \\\cline{1-2}
\rule{0pt}{4ex} $X_2$ & $A_2+A_4+A_5$ \\\cline{1-2}
\end{tabular}
\label{feynrulesTabla3}
\end{table}

\begin{table}[H]
\centering
\caption{Factors from Table \ref{feynrulesTabla}.}
\begin{tabular}{|c|c|}\cline{1-2}
\rule{0pt}{4ex} $B_1$ & $y_1y_2(y_1^2+y_3^2)c_{\beta}^2$ \\\cline{1-2}
\rule{0pt}{4ex} $B_2$ & $y_1^3y_2s_{\beta}-\dfrac{1}{2}y_1y_2y_3^2s_{\beta}$ \\\cline{1-2}
\rule{0pt}{4ex} $B_3$ & $\dfrac{fg}{4m_W}y_3(2y_1^2+5y_2^2)\sqrt{y_1^2+y_3^2}s_{\beta}$ \\\cline{1-2}
\rule{0pt}{4ex} $B_4$ & $-y_1^3y_3s_{\beta}-\dfrac{1}{2}y_1(y_2^2+y_3^2)s_{\beta}$ \\\cline{1-2}
\rule{0pt}{4ex} $A_1$ & $\dfrac{-4m_Wc_{\beta}^2y_1y_2(y_1^2+y_3^2)}{s_{\beta}y_3(2y_1^2-y_2^2)}$ \\\cline{1-2}
\rule{0pt}{4ex} \multirow{2}{*}{$A_2$} & $fg-y_1^8-m_Ws_{\beta}y_3y_2^6-3y_2y_1^4(2y_1^2-y_2^2)$ \\
\rule{0pt}{4ex} & $-y_1^6(3y_2^2+y_3^2)-y_1^2y_2^4(y_2^2+3y_3^2)$ \\\cline{1-2}
\rule{0pt}{4ex} $A_3$ & $\dfrac{-4m_Wy_1^3y_2y_3}{2y_1^2-y_2^2}$ \\\cline{1-2}
\rule{0pt}{4ex} $A_4$ & $\dfrac{8m_Wy_1^3(y_2^2+y_3^2)}{2y_1^2-y_2^2}$ \\\cline{1-2}
\rule{0pt}{4ex} $A_5$ & $\dfrac{-4m_Wy_1^3}{2y_1^2-y_2^2}$ \\\cline{1-2}
\rule{0pt}{4ex} $A_6$ & $\dfrac{4m_Wy_1^3y_2}{y_3(2y_1^2-y_2^2)}$ \\\cline{1-2}
\rule{0pt}{4ex} $\Delta_1$ & $p_{\alpha}-q_{\alpha}$ \\\cline{1-2}
\rule{0pt}{4ex} $\Delta_2$ & $q_{\beta}-k_{\beta}$ \\\cline{1-2}
\rule{0pt}{4ex} $\Delta_3$ & $k_{\mu}-p_{\mu}$ \\\cline{1-2}
\end{tabular}
\label{feynrulesTabla4}
\end{table}

\end{document}